\pdfoutput=1

\documentclass[11pt]{article}

\usepackage[preprint]{acl}

\usepackage{times}
\usepackage{latexsym}

\usepackage{hyperref}
\usepackage[T1]{fontenc}

\usepackage[utf8]{inputenc}

\usepackage{microtype}

\usepackage{inconsolata}

\usepackage{pifont}

\usepackage{graphicx}
\usepackage{multirow}
\usepackage{booktabs}
\usepackage{amsmath}

\title{A Survey of Generative Information Retrieval}

\author{Tzu-Lin Kuo$^*$\quad Tzu-Wei Chiu$^*$\quad Tzung-Sheng Lin$^*$\\
\textbf{Sheng-Yang Wu$^*$\quad Chao-Wei Huang\quad Yun-Nung Chen}\\
  National Taiwan University, Taipei, Taiwan \\
  \texttt{\{r12922050,r12922a14,r10922a24,r12944033\}@ntu.edu.tw}\\
  \texttt{f07922069@csie.ntu.edu.tw\quad y.v.chen@ieee.org}\\
}

\begin{document}
\maketitle
\begin{abstract}
Generative Retrieval (GR) is an emerging paradigm in information retrieval that leverages generative models to directly map queries to relevant document identifiers (DocIDs) without the need for traditional query processing or document reranking. This survey provides a comprehensive overview of GR, highlighting key developments, indexing and retrieval strategies, and challenges. We discuss various document identifier strategies, including numerical and string-based identifiers, and explore different document representation methods. Our primary contribution lies in outlining future research directions that could profoundly impact the field: improving the quality of query generation, exploring learnable document identifiers, enhancing scalability, and integrating GR with multi-task learning frameworks. By examining state-of-the-art GR techniques and their applications, this survey aims to provide a foundational understanding of GR and inspire further innovations in this transformative approach to information retrieval. We also make the complementary materials such as paper collection publicly available\footnote{\url{https://github.com/MiuLab/GenIR-Survey/}}.
\begingroup\def\thefootnote{\rm *}\footnotetext{Equal contribution.}\endgroup
\end{abstract}

\section{Introduction}
\begin{figure*}[t]
  \centering
  \includegraphics[width=\textwidth]{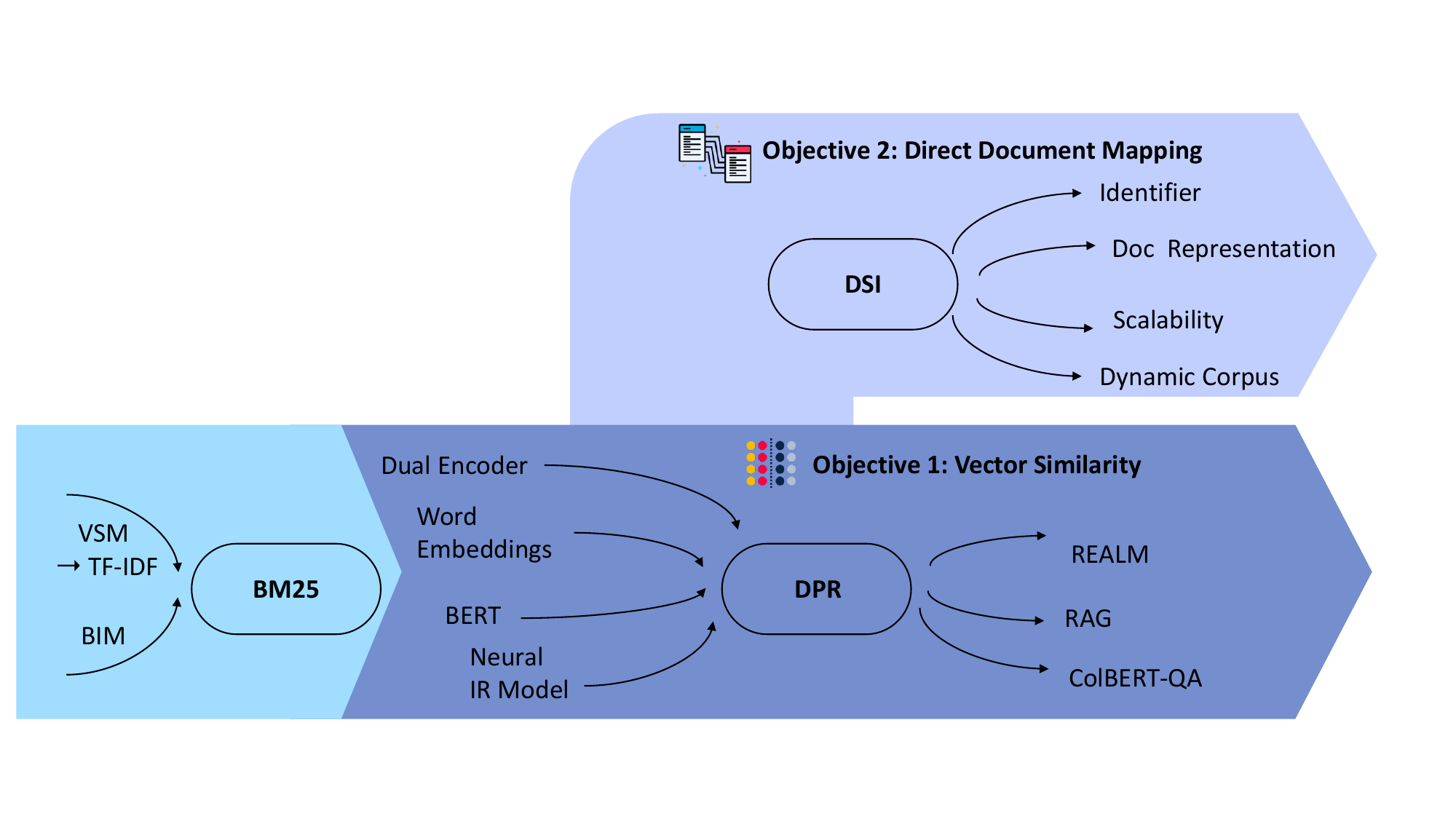}
  \caption{Progression of information retrieval from sparse vector similarity techniques, such as the bag-of-words and Vector Space Model, to dense retrieval with innovations like Word2Vec and BERT, culminating in sophisticated systems like DPR. Advances in generative retrieval now integrate language models for direct response generation.}
  \label{fig:Objectives}
\end{figure*}

The history of Information Retrieval (IR) has undergone significant evolution, transitioning from rudimentary methods grounded in statistical word relationships to sophisticated systems that leverage advanced deep learning techniques. This progression is distinctly organized around two primary training objectives, as illustrated in Figure \ref{fig:Objectives}:

\paragraph{Objective 1: Vector Similarity} Initially, IR systems were dependent on sparse retrieval techniques that utilized statistical relationships between words through methods such as the bag-of-words approach and the Vector Space Model (VSM) \cite{salton1983introduction}. In these models, documents were represented as sparse vectors, with each dimension indicating the presence or frequency of terms. The development of the Binary Independence Model (BIM) \cite{robertson1976relevance} and the implementation of Term Frequency-Inverse Document Frequency (TF-IDF) are quintessential to this approach, emphasizing the independence and frequency of term occurrences.

As technological advancements emerged, the emphasis shifted towards dense retrieval. In this phase, word embeddings transformed words into dense vector representations, capturing deeper semantic similarities and contextual relationships beyond mere keyword matches. Prominent developments in this area include Word2Vec \cite{mikolov2013efficient}, GloVe \cite{pennington2014glove}, and advances in transformer networks such as BERT \cite{devlin2018bert}. These innovations culminated in sophisticated models like Dense Passage Retrieval (DPR) \cite{karpukhin2020dense}, which markedly enhanced the precision and effectiveness of information retrieval by employing dense vector embeddings to comprehend complex queries and documents.
Building on DPR, models like REALM \cite{guu2020retrieval} and RAG \cite{lewis2020retrieval} integrate retrieval with language models, further refining relevance. ColBERT-QA \cite{khattab2021relevance} employs contextualized embeddings for precise answer retrieval, advancing question-answering capabilities.

\paragraph{Objective 2: Direct Document Mapping} As IR has transitioned from vector similarity approaches, it has embraced generative retrieval, a method that employs generative models to directly produce text responses or document identifiers relevant to user queries. This marks a significant shift from matching pre-existing vector representations to dynamically generating textual outputs that directly address user needs. In the pre-retrieval stage, generative models are utilized to enhance dense retrieval efficiency through innovative approaches like a retrieval-oriented pre-training paradigm using a Masked Auto-Encoder (MAE), as demonstrated by \citet{xiao2022retromae}. This model trains to reconstruct sentences from their embeddings and masked inputs, delivering superior performance across various benchmarks. During the retrieval phase, the adoption of generative models is exemplified by \citet{lewis2020retrieval}'s retrieval-augmented generation model, which selects documents via a dense passage retriever and generates answers for complex NLP tasks, achieving top-tier performance. Moreover, \citet{tay2022transformer}'s Differentiable Search Index (DSI) highlights this stage by mapping queries directly to relevant documents, significantly surpassing traditional methods and demonstrating robust generalization in zero-shot setups. In the post-retrieval phase, deep learning techniques are applied to rerank retrieved documents, where efforts like those by \citet{guo2016deep} focus on refining document rankings by analyzing complex matching patterns between queries and documents. Similarly, \citet{mitra2017learning} enhances web search reranking by merging local and distributed text representations, leveraging both local and global contexts to improve search result quality. Through these innovations, including the Two-Tower model architecture and the Differentiable Search Index (DSI) \cite{tay2022transformer}, generative retrieval not only effectively responds to queries but also identifies pertinent information within a corpus, utilizing end-to-end training architectures that integrate deep learning processes to streamline the retrieval experience.

\section{Introduction to Generative Retrieval}
\subsection{Definition of Generative Retrieval}
\begin{figure*}
    \centering
    \includegraphics[width=1\linewidth]{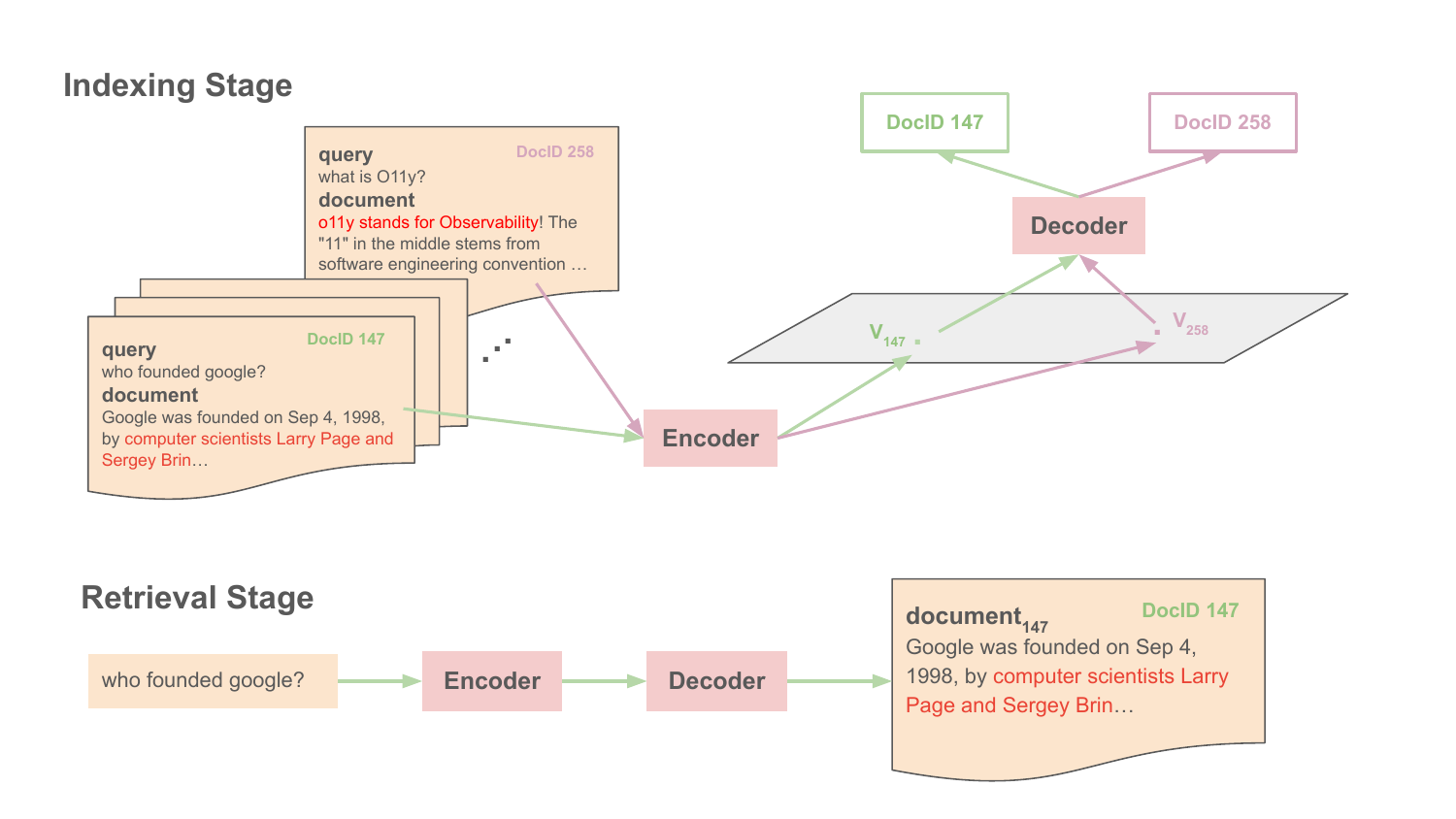}
    \caption{The Generative Retrieval system consists of two primary stages: In the Indexing Stage, specific queries like "What is O11y?" and "Who founded Google?" are linked with their corresponding documents to DocIDs (DocIDs 258 and 147, respectively) through a seq2seq learning system, ensuring accurate query-DocID and document-DocID associations. The Retrieval Stage processes a user query ("Who founded Google?") to autoregressively output the relevant DocID, eliminating the need for additional query processing and document reranking. This direct mapping highlights the system's capability for efficient, end-to-end retrieval based on learned relationships.}
    \label{fig:gr-pipeline-indexing}
\end{figure*}

The preceding section demonstrates the application of generative models at various IR stages to facilitate task execution. In this survey paper, we aim to define "generative retrieval" (GR) in the context of the Differentiable Search Index architecture \cite{tay2022transformer}, wherein a query is directly mapped to its relevant document(s) through a seq2seq model without the need for pre-retrieval query processing or post-retrieval reranking of documents. Essentially, an end-to-end architecture is sufficient for completing IR tasks. Formally, we define GR as a system where, given a user query $q$ as input, a seq2seq learning model directly outputs several document identifiers (docids). Each identifier $j$ corresponds to a specific document $d_j$ within a corpus $D$, indicating that the document is relevant to query $q$ (See Figure \ref{fig:gr-pipeline-indexing}). Achieving this requires two critical components in GR: indexing and retrieval.

\subsubsection{Indexing}
\label{sec:intro-to-gr-indexing}
In the GR indexing strategy, pivotal considerations are \textbf{indexing methods} and \textbf{indexing targets}. Indexing methods examines the techniques used to establish the connection between document content and their unique identifiers, essentially mastering the process of associating each document's text with a distinct docid. Indexing targets, conversely, focuses on document representation strategies. This involves decisions regarding the detail level of indexing, the importance of indexing specific document sections over others, handling duplicate information, and the importance of semantic comprehension in depicting the essence of document contents.

In GR's indexing methods, the emphasis is on streamlining the process that connects document contents with their unique identifiers. We can formulate the process of indexing methods as training on two types of examples. The first one is $(d_j, j)$ where $d_j \in D$ represents $j$-th document in the corpus $D$ and $j$ represents the corresponding identifier. It is essential to training on the document-docid pairings for building the index. This pairing process is the first step in creating a retrievable link between the content of each document and its location within the database, enabling efficient storage and retrieval.

And the second training example is $(q_i, j)$, where we link a query $q_i$ with its relevant docid $j$. By pairing queries with relevant docids, the system learns the contextual nuances that define relevance between a user's search intent (expressed through the query) and the document content (represented by the docid). This training helps the model understand which documents are most relevant to a given query, an understanding that cannot be achieved by indexing alone. These methods include innovative approaches to sequence-to-sequence conversions and bidirectional training, as well as advanced techniques for span-based denoising. 
The details in the second training example will be covered in Section \ref{sec:docid}.

For indexing targets, the focus shifts to how documents are represented within the system. Due to limitations in model capacity and computational resources, it's often impractical for generative retrieval models to train with entire documents as direct inputs. Therefore, it's necessary to consider other methods to effectively represent documents, which include:

1. Direct Indexing: take the first L tokens of the document.

2. Set Indexing: take the first L tokens without repeated words.

3. Inverted Index: take k contiguous tokens starting randomly from the document.

4. Queries as representation: \citet{zhuang2022bridging} proposed a method that uses generated queries to represent documents while training with the DocID. They suggested that using queries for training instead of the whole document aligns better with the retrieval process, as retrieval typically involves using queries to find relevant documents.

By employing these diverse indexing methods, we aim to enhance the efficiency and accuracy of the generative retrieval system. Direct Indexing and Set Indexing offer straightforward yet effective means to capture essential document content while minimizing redundancy. The Inverted Index provides a randomized, yet systematic, approach to document representation, ensuring diverse content coverage. Meanwhile, leveraging queries as document representations aligns the training phase with the retrieval phase, promoting a more intuitive and context-aware retrieval process.

Ultimately, these indexing strategies converge towards a unified goal: to optimize the generative retrieval system's ability to understand, index, and retrieve documents with high precision. By balancing detail, relevance, and comprehensiveness, we can ensure that the system not only stores document content efficiently but also retrieves the most pertinent information accurately in response to user queries. This balance is crucial for developing a robust and scalable generative retrieval framework capable of handling diverse and complex information needs.

\subsubsection{Retrieval}

Upon completing the indexing phase, we turn our attention to the retrieval phase. Classic GR model adopts a seq2seq approach for autoregressively decoding candidate docids, where the choice of representation for these docids critically affects retrieval efficiency. 

In the seminal work of Generative Retrieval, \citet{tay2022transformer} introduces Unstructured Atomic Identifier method, which assigns unique integers to each document. This foundational approach is complemented by structured identifier methods, including both naively structured string identifiers and semantically structured identifiers, paving the way for nuanced document representation. As the field has evolved, subsequent works have diversified the focus on identifier representation, exploring alternatives like subsets of strings, article titles, and more. Section 3 will provide a detailed exploration and comparison of these extensions and the broader spectrum of works within this series, highlighting their contributions and innovations in the context of Generative Retrieval.

\section{Different Document Identifier Strategies}
\label{sec:docid}

\begin{figure*}
    \centering
    \includegraphics[width=0.75\linewidth]{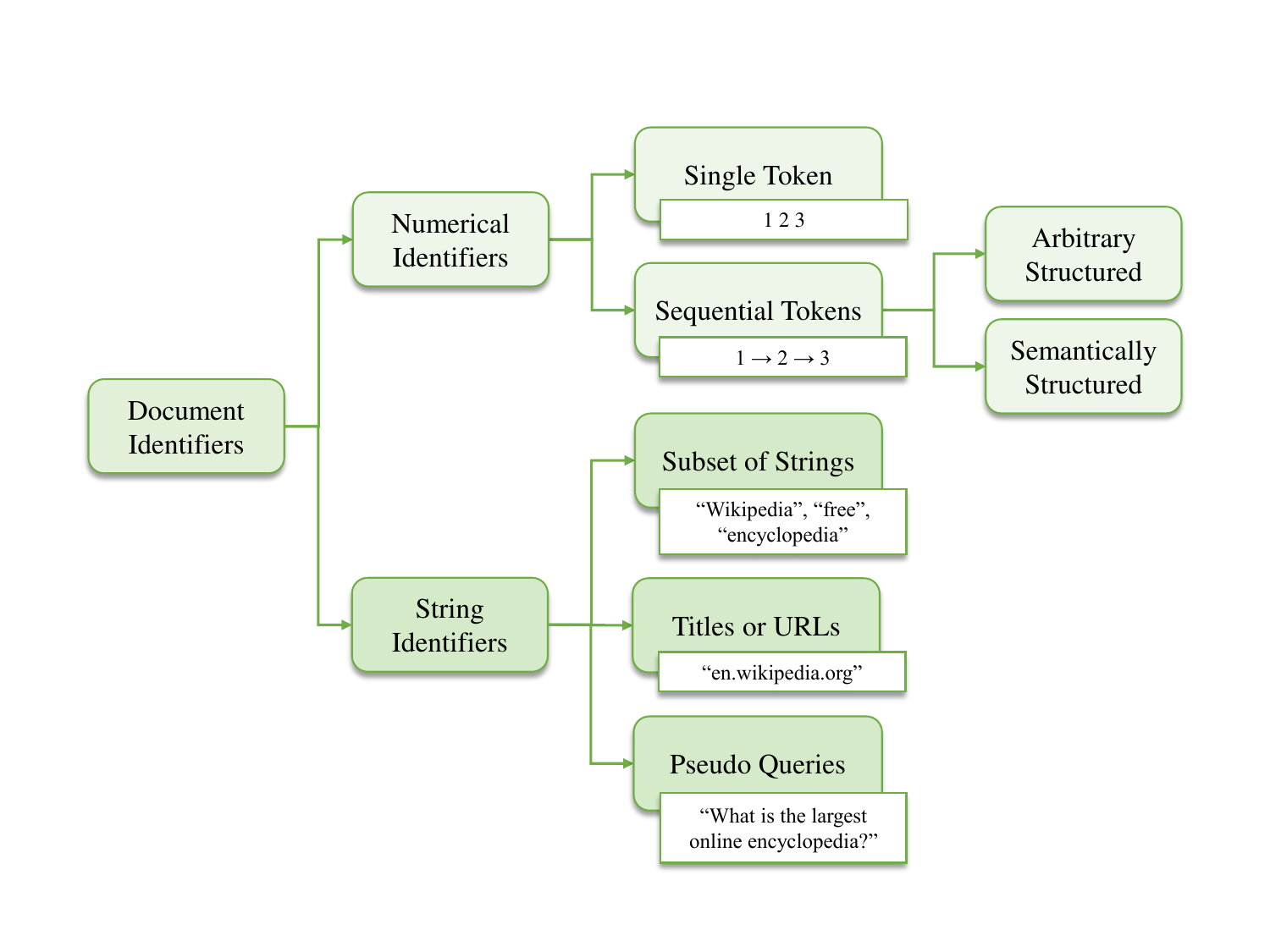}
    \caption{
        Different types of document identifiers. We categorize docid into two types: numerical identifiers and string identifiers. Numerical identifiers use numbers as identifiers and are further classified into single token and sequential tokens based on the number of tokens used to represent each docid. With sequential tokens, the model decodes tokens sequentially, one by one, for each docid. Depending on the method used to create the hierarchy structure, sequential tokens can be further divided into arbitrary structured and semantically structured identifiers. With string identifiers, the model directly decodes strings as docid. Based on the type of string used, we divide them into subset of strings, titles or URLs, and pseudo queries. 
    }
    \label{fig:docid}
\end{figure*}

Why is document identitier (docid) so important for Generative Retrieval? According to \citet{tang2023recent}, instead of letting the generative model generate the entire document by memorizing, it is easier to only memorize and generate the document identifiers. Therefore, it is important to have a well-designed document identifier than not only represent the document well but also have a good performance at retrieving each relevant document.

In the following sections, we will roughly divide the type of document identifiers into two main types: 1. numerical docid and 2. string docid, which the generative model either generates either numerical token(s) or strings to identify different documents. The structure used to categorise document identifiers' type is shown in Figure \ref{fig:docid}. Full table of the list of works according to the type of document identifiers are shown in Table \ref{docid_works}. We also listed the methods that are used to create document identifiers in Table \ref{docid_methods}.

\subsection{Numerical Identifier}

Numerical Document Identifiers (docids) uses tokens or sequence of tokens to represent each document.  We can categorize numerical docids by the number of tokens used to identify each document.

1. Single Token: This is a simple method that assigns each document an arbitrary unique token for generative models, such as T5's final layer to decode. Each identifier can be a random integer or class label that is unique. The generative model decodes each document using only one token. Examples of works that use single token identifiers include those by \citet{tay2022transformer, mehta2022dsi++, zhou2022dynamicretriever, nguyen2023generative, chen2023continual}

2. Sequential Tokens: This approach uses unique sequence of tokens as identifier. The generative model decodes tokens sequentially one token at a time. We can further divide it into two categories by the relations between the sequence of tokens. 

The first category uses arbitrary sequence of tokens, which can be random initialized. This means that there aren't relationship between the hierarchical of the sequence that the model is decoding, identifiers with same sub-sequence of tokens can be completely irrelevant. Examples of works with arbitrarily structured sequences of tokens include those by \citet{tay2022transformer, mehta2022dsi++, zhuang2022bridging, nadeem2022codedsi}.

The second category use semantically structured sequence of tokens, which same starting sub-sequence means the documents contain more similar semantic information. There are two ways to construct semantically structured sequences in the works we surveyed. The first method, proposed in DSI \cite{tay2022transformer}, captures semantic information by using K-means clustering similar documents together with a hierarchical structure. The generative model then retrieves sequentially to find the document within the cluster of similar documents. This method is more commonly seen in the works we found. Examples of works using clustering methods include those by \citet{tay2022transformer, mehta2022dsi++, nadeem2022codedsi, nguyen2023generative, wang2022neural, zhou2022ultron}. Another method from prior work, proposed by \citet{sun2024learning}, uses an autoencoder architecture, which captures the semantic information by encoding the original document to docid and tries to regenerate the same document from decoding the docid. The retrieving process is same as the first method, by sequentially docoding each discrete identifier.

The performance varies among different types of numerical document identifiers used. More studies suggest that methods with semantic information performs better. However, \citet{mehta2022dsi++} indicated that using semantic structure identifiers reduces forgetting, but still underperforms unstructured document identifiers.

\begin{table*}
\centering\small
\begin{tabular}{|l|ccc|ccc|}
\hline
\multirow{2}{*}{\textbf{Methods}} & \multicolumn{3}{c|}{\textbf{Numerical Identifiers}} & \multicolumn{3}{c|}{\textbf{String Identifiers}} \\
{} & \textbf{single} & \textbf{sequence} & \textbf{w/semantic} & \textbf{string sets} & \textbf{titles/URLs} & \textbf{queries} \\
\hline\hline

DSI \cite{tay2022transformer} & \ding{51} & \ding{51} & \ding{51} & & & \\ \hline
DSI++ \cite{mehta2022dsi++} & \ding{51} & \ding{51} & \ding{51} & & & \\ \hline
CodeDSI \cite{nadeem2022codedsi} & & \ding{51} & \ding{51} & & & \\ \hline
DynamicRetriever \cite{zhou2022dynamicretriever} & \ding{51} & & & & & \\ \hline
Tied-Atomic \cite{nguyen2023generative} & \ding{51} & & \ding{51} & & & \\ \hline
NCI \cite{wang2022neural} & & & \ding{51} & & & \\ \hline
DSI-QG \cite{zhuang2022bridging} & & \ding{51} & & & & \\ \hline
CELVER \cite{chen2023continual} & \ding{51} & & & & & \\ \hline
Ultron \cite{zhou2022ultron} & \ding{51} & & \ding{51} & & \ding{51} & \\ \hline
GENRET \cite{sun2024learning} & & & \ding{51} & & & \\ \hline
GERE \cite{chen2022gere} & & & & & \ding{51} & \\ \hline
CorpusBrain \cite{chen2022corpusbrain} & & & & & \ding{51} & \\ \hline
DEARDR \cite{thorne2022data} & & & & & \ding{51} & \\ \hline
GENRE \cite{de2020autoregressive} & & & & & \ding{51} & \\ \hline
SEAL \cite{bevilacqua2022autoregressive} & & & & \ding{51} & & \\ \hline
TSGen \cite{zhang2023term} & & & & \ding{51} & & \\ \hline
SE-DSI \cite{tang2023semantic} & & & & & & \ding{51} \\ \hline
TOME \cite{ren2023tome} & & & & & \ding{51} & \\ \hline
MINDER \cite{li2023multiview} & & & & \ding{51} & \ding{51} & \ding{51} \\ \hline
LTRGR \cite{li2024learning} & & & & \ding{51} & \ding{51} & \ding{51} \\ \hline

\end{tabular}
\caption{\label{docid_works}
List of works and types of docids used. W/semantic means semantically structured sequential tokens.
}
\end{table*}

\begin{table*}
\centering
\small
\begin{tabular}{llcc}
\hline
\textbf{Method} & \textbf{Description} & \textbf{Identifier type} &
\textbf{Semantic information} \\
\hline
Arbitrary & random initialized token(s) & numerical &  \\
Clustering & use K-means clustering & numerical & \ding{51} \\
Auto-encoder & encode document to embedding & numerical & \ding{51} \\
String set & n-grams, substrings, URLs or titles & string & \ding{51} \\
\hline
\end{tabular}
\caption{\label{docid_methods}
Methods for creating document identifiers
}
\end{table*}

\subsection{String Identifier}

Different from numerical docid, String Identifiers use multiple words as docids, which typically contain semantic information. Based on the type of string they use, we can categorize them into three categories. 

1. Subset of strings: This method uses a subset of strings from the document as the document identifier. There are various ways to obtain the subset of strings. We list two methods that use this strategy: n-gram and unordered term set. \citet{bevilacqua2022autoregressive} used a conditioned generative model to generate n-grams that exist in the documents to retrieve the relevant document. \citet{zhang2023term} proposed a method that use N unordered unique terms as document identifier, where they stated that any permutations of the terms should retrieve the document. 

2. Article Titles or URLs: This approach uses the titles or URLs of the documents as identifiers, which usually contain some semantic information about the document. The generative model directly generates the tokenized document title or URLs when retrieving. Examples of works that used article titles as identifiers include those by \citet{chen2022gere, de2020autoregressive, chen2022corpusbrain, thorne2022data}, and works that used URLs as identifiers include \citet{zhou2022ultron, ren2023tome}.

3. Synthetic pseudo-queries: \citet{tang2023semantic} utilized a generation model (DocTTTTTQuery 
 by \citet{nogueira2019document}) to generate pseudo-queries of each given document, and used the top 1 generated query as the document identifier for retrieval. When retrieving, the rank list is generated by beam search. There is some portion of the identifiers that is not unique. To deal with the problem, they used a random order for documents with the same docid during inference time. 

Additionally, there are some other methods that use more than one strategy at once. The method proposed by \citet{li2023multiview} introduced a multi-view identifier that combines the methods above. The multi-view identifier contains titles, substrings, and pseudo-queries together to better identify the document. Prior work by \citet{li2024learning} also used the same type of identifier. 

As mentioned by \citet{ren2023tome}, one advantage of using string type of identifiers is that these type of identifiers are more aligned to the tokens used to pretrain the language model.

Methods with using identifiers that contains semantic information are often regarded as the better option in more situations from the works we found.
The process of creating arbitrary docids are easier, but the performances are not that consistently reliable.

\section{Evaluation of Generative Information Retrieval}
\subsection{Performance Metrics}

Our study primarily investigates the performance differences among various generative retrievals. We reference the experimental outcomes of BM25 and T5-based dual encoders to establish baselines for sparse and dense retrieval methods. In generative retrieval research, common evaluation metrics include Hits, Recall, and Mean Reciprocal Rank (MRR). For studies with the objective of Direct Document Mapping, the Differentiable Search Index (DSI) \cite{tay2022transformer} is one of the earlier approaches and serves as a comparative benchmark in many generative retrieval studies. The primary metrics used in these comparisons are Hits@1 and Hits@10. Hits@N, where N={1, 10}, reports the proportion of correct documents ranked within the top N predictions.

\subsection{Datasets}

We compile a comprehensive list of datasets used in various generative retrieval experiments in table \ref{tab:datasets}. The datasets include MS MARCO \cite{nguyen2016ms}, Natural Questions \cite{kwiatkowski2019natural}, Trivia QA \cite{joshi2017triviaqa}, FEVER \cite{thorne2018fever}, codeSearchNet \cite{husain2019codesearchnet}, XOR QA \cite{asai2020xor}, KILT \cite{petroni2020kilt}, and HOVER \cite{jiang2020hover}. We evaluate the relative performance of these methods by selecting studies that utilize the two most commonly employed datasets, as determined by their frequency of use in the literature. This approach allows for a detailed comparative analysis based on widely recognized benchmarks.

\textbf{MS MARCO} \cite{nguyen2016ms} includes 1,010,916 questions from Bing with human-generated answers, 182,669 rewritten answers, and 8,841,823 passages from web documents. Generative retrieval studies use varying data sizes (10k, 100k, 300k, 1M, full) for different experiments. Table \ref{tab:msdoc-res} compares models with the same data size and base model.

\textbf{Natural Questions (NQ)} \cite{kwiatkowski2019natural} includes real Google queries with annotated answers from Wikipedia. Generative retrieval studies often use 10k, 100k, or 320k examples. Table \ref{tab:nqdoc-res} compares models using the most frequent 320k dataset with the same base model.

\begin{table}
    \centering
     \renewcommand{\arraystretch}{1.2}%
    \small
    \begin{tabular}{l}
    \hline
       \textbf{Datasets} \\
    \hline
       MS MARCO$^\dag$ \cite{nguyen2016ms} \\
       Natural Questions (NQ)$^\dag$ \cite{kwiatkowski2019natural} \\
       Trivia QA \cite{joshi2017triviaqa} \\
       FEVER \cite{thorne2018fever} \\
       codeSearchNet \cite{husain2019codesearchnet} \\
       XOR QA \cite{asai2020xor} \\
       KILT \cite{petroni2020kilt} \\
       HOVER \cite{jiang2020hover} \\
    \hline
    \end{tabular}
    \caption{List of datasets used. $^\dag$ indicates commonly used datasets.}
    \label{tab:datasets}
\end{table}

\subsection{Baselines and Comparisons}
\label{baseline-comparison}
The performance of various methods for generative retrieval tasks is illustrated through two tables, evaluating the models on different datasets: MS MARCO and NQ. In the MS MARCO dataset, several data sizes are considered, including 100K, 300K, and the full dataset. For the NQ dataset, a single data size of 320K is evaluated.

In Table~\ref{tab:msdoc-res}, the MS MARCO dataset serves as a benchmark to evaluate various methods including TOME (both single-stage and two-stage), SE-DSI, TSGen, Tied-Atomic, GENRET, and Ultron, each employing the T5-base model \cite{raffel2020exploring} as their underlying architecture. Within the MS MARCO 100K subset, the TOME$_{\text{two-stage}}$ method achieves a leading Hits@1 score of 71.93, whereas SE-DSI records the highest Hits@10 score of 75.28. In the context of the MS MARCO 300K subset, TSGen outperforms others in both Hits@1, registering a score of 38.40, and Hits@10, with a score of 78.10. Across the entire MS MARCO dataset, SE-DSI secures the top Hits@1 score of 26.09 and the highest Hits@10 score of 40.02. The BM25 method is also included as a baseline for comparison but exhibits lesser performance.

Table~\ref{tab:nqdoc-res} focuses on the NQ 320K dataset, comparing methods such as DSI, Tied-Atomic, NCI, NCI$_{w/qq-ft}$, Ultron, GENRET, TSGen, DSI-QG, and TOME, all using T5-base \cite{raffel2020exploring} as their base model. Tied-Atomic, employing a single document identifier, achieves state-of-the-art performance with the highest Hits@10 score of 90.03. On the other hand, TSGen, which utilizes string set document identifiers, records the highest Hits@1 score of 70.80. Overall, the document identifier strategy employing string representations demonstrates superior performance in the Hits@1 metric. However, TSGen does not exhibit the best performance on the MS MARCO 300K dataset. This variation in results between datasets suggests inherent differences in their characteristics. Consequently, no single method consistently outperforms across different datasets. In contrast, the BM25 baseline method shows significantly lower scores, with 15.11 for Hits@1 and 32.48 for Hits@10, underscoring the advanced effectiveness of the newer models.

In addition to studies using T5-base as the base model, there is a segment of research that uses BART \cite{lewis2019bart} as the base model. Notable examples include SEAL \cite{bevilacqua2022autoregressive}, MINDER \cite{li2023multiview}, and LTRGR \cite{li2024learning}, all of which use BART-large to conduct experiments on the MS MARCO and NQ datasets. These studies also employ common methodologies, such as the use of string identifiers for comparative analysis. Specifically, in the NQ 320K dataset,  LTRGR exhibits state-of-the-art performance, achieving the highest results in metrics like hit@5, hit@20, and hit@100. Given the relatively sparse volume of research and experiments of this nature, these findings are discussed textually rather than presented in table format, highlighting their significance and the unique contributions of using BART-large as a base model in information retrieval tasks.

Overall, the tables highlight the superior performance of specific generative retrieval methods like TOME$_\text{two-stage}$, SE-DSI, TSGen, and Tied-Atomic under various experimental conditions and dataset sizes. The best results for Hits@1 and Hits@10 are highlighted in bold, showcasing the effectiveness of these models compared to the baseline BM25.

\subsection{Document Identifier Strategy Comparison}
From Table~\ref{tab:msdoc-res} and Table~\ref{tab:nqdoc-res}, we can observe that docids with semantic information usually have better performance than the docids without semantic information. 

However, in Ultron \cite{zhou2022ultron}, the best performance comes from the atomic setting, which is the single token numerical docid. They addressed that the reason it performs better is that it used almost about twice the parameter size of the semantic ones such as the URLs. This is also one of the problems of single token numerical docid. With the size of the corpus increasing, the model's final output layer needs to match the number of the documents of the whole corpus size, which leads to an increase in parameters of the model. On the other hand, semantic docids such as sequential tokens and string identifiers have limited amount of tokens. The model only needs to output the number of groups for each hierarchy, such as k for the k-means clustering, or the limited amount of strings for URLs and titles which tokens are limited to the amount of subwords that don't scale with the corpus.  In this case, we think that with increasing the parameters of the semantic docid models will leads to better performance. 

In the case above, we can observe that although the tables compare different works with the same dataset size and the same base model, the difference of model parameter size leads to different performance. Therefore, more thorough experiments with the same settings might be needed for a more in-depth comparison. 

\begin{table*}
    \small
    \centering
    \setlength{\tabcolsep}{3pt}%
    \renewcommand{\arraystretch}{0.95}%
    \begin{tabular}{l|c|cccccccccccccccc}
        \toprule
        \multirow{2}{*}{\textbf{Methods}} & \multirow{2}{*}{\textbf{docid}}  &  \multicolumn{2}{c}{\textbf{MS MARCO 100K}} &  \multicolumn{2}{c}{\textbf{MS MARCO 300K}} &\multicolumn{2}{c}{\textbf{MS MARCO Full}}\\
        & & \textbf{Hits@1} & \textbf{Hits@10}  & \textbf{Hits@1} & \textbf{Hits@10} & \textbf{Hits@1} & \textbf{Hits@10} \\
        \midrule

        $\textbf{BM25}$ & - & - &  - &  - &  - & - & 18.92\\
        \midrule
        $\textbf{TOME}_{single-stage}$  & URL & 66.46 &  - &  - &  - & - & - \\
        $\textbf{TOME}_{two-stage}$  & URL & \textbf{71.93} &  - &  - &  - & 22.03 & - \\
        $\textbf{SE-DSI}$  & query & 53.47 &  75.28 &  - &  - & \textbf{26.09} & \textbf{40.02} \\
        $\textbf{TSGen}$  & str set & - &  - &  38.40 &  78.10 & - & -  \\
        $\textbf{GENRET}$ & w/sem & - &  - &  \textbf{47.90} &  \textbf{79.80} & - & - \\
        $\textbf{Ultron}_{atomic}$ & single & - &  - &  32.81 &  74.13 & - & - \\

        \bottomrule
    \end{tabular}
    \caption{Comparison of generative retrieval experiments using varying data sizes (100k, 300k, 1M, full) with T5-base \cite{raffel2020exploring} as the base model. Despite using the same base model, the data sizes and experimental settings differ slightly. The best results for Hits@1 and Hits@10 under similar experimental settings are highlighted in bold. Note that the experiment results are directly sourced from their respective papers. Categories of docids can be found in Table \ref{docid_works}, with seq referring to sequence, str referring to string and w/sem referring to semantically structured sequential tokens. }
    \label{tab:msdoc-res}
    \vspace*{-2mm}
\end{table*}

\begin{table}
    \small
    \centering
    \setlength{\tabcolsep}{3pt}%
    \renewcommand{\arraystretch}{0.95}%
    \begin{tabular}{l|c|cc}
        \toprule
        \multirow{2}{*}{\textbf{Methods}} & \multirow{2}{*}{\textbf{docid}} &  \multicolumn{2}{c}{\textbf{NQ 320K}} \\
        & & \textbf{Hits@1} & \textbf{Hits@10}  \\
        \midrule
        $\textbf{BM25}$ & - & 11.60 &  34.40 \\
        \midrule
        $\textbf{T5-based dual encoders}$ & - & 20.50 &  58.30 \\
        \midrule
        $\textbf{DSI}$  & w/sem & 27.40 &  56.60  \\
        $\textbf{Tied-Atomic}$  & single & 65.26 &  \textbf{90.03}  \\
        $\textbf{NCI}$  & w/sem & 65.86 &  85.20  \\
        $\textbf{NCI}_{w/ qg-ft}$  & w/sem & 68.91 &  88.48  \\
        $\textbf{Ultron}_{atomic}$  & single & 25.43 &  69.53  \\
        $\textbf{GENRET}$  & w/sem & 68.10 &  88.80  \\
        $\textbf{TSGen}$ & str set & \textbf{70.80} & 88.90 \\
        $\textbf{DSI-QG}$ & seq & 63.49 &  82.36  \\
        $\textbf{TOME}_{single-stage}$ & URL & 64.93 &  -  \\
        $\textbf{TOME}_{two-stage}$ & URL & 66.64 &  -  \\

        \bottomrule
    \end{tabular}
    \caption{Comparison of generative retrieval experiments using 10k, 100k, and 320k examples with T5-base \cite{raffel2020exploring} as the base model. The majority of studies utilize the 320k dataset. The best results for Hits@1 and Hits@10 under similar experimental settings are highlighted in bold. Note that the experiment results are directly sourced from their respective papers. The experimental data for BM25 and T5-based dual encoders is sourced from the DSI authors.  Settings w/qg-ft in NCI refers to query generation. Categories of docids can be found in Table \ref{docid_works}, with seq referring to sequence, str referring to string and w/sem referring to semantically structured sequential tokens. }
    \label{tab:nqdoc-res}
    \vspace*{-2mm}
\end{table}

\section{Challenges}
Although generative retrieval appears to be an innovative direction and has shown outstanding performance in various experiments, there are still some issues that need to be addressed. These challenges include the problem of scalability, which involves maintaining efficient retrieval performance on large-scale datasets, and the problem of dynamic corpora, which requires the system to dynamically handle and update the content in the database to ensure the accuracy and timeliness of retrieval results. Only by overcoming these challenges can generative retrieval realize its full potential.

\subsection{Scalability}

According to the experimental results by \citet{tay2022transformer}, it was found that the performance improvement of the Generative Retrieval Model decreased as the dataset size expanded from 100k to 320k. Generally, most studies on generative retrieval focus on evaluating datasets around the size of 100k, proving the model's strong competitiveness at this scale. However, in real-world applications of information retrieval, the dataset sizes often far exceed this magnitude.

Therefore, \citet{pradeep2023does} conducted an in-depth study on the entire MS MARCO dataset, which contains over 8.8 million paragraphs, to observe how the performance of generative retrieval model is affected as the dataset size increases. Notably, in the training dataset of MS MARCO, less than 6\% of the documents contain queries, which is closer to the real-world situation where most documents do not have a user query record.

As the dataset expanded from MSMarco100k to MSMarcoFULL, the performance gap between the Generative Retrieval Model and the baseline model in the MRR@10 metric gradually widened, increasing from 2.9 to 10.2. This indicates that the increase in dataset size poses a greater challenge for the Generative Retrieval Model, showing that its performance is not only inferior to the baseline model but also that this performance gap continues to grow as the dataset size increases.

Furthermore, \citet{pradeep2023does} also analyzed techniques previously applied in generative retrieval models to determine which techniques could effectively mitigate the challenges brought about by the increase in dataset size. The study found that the training method using synthetic queries to document identifiers (docid) exhibited the best performance on large-scale data, proving the effectiveness of this approach. Additionally, the quality of synthetic queries also significantly impacted the model's performance. This training method is effective mainly because it increases the proportion of documents in the dataset that contain queries, thereby reducing the discrepancy between training and inference stages when using documents and queries as inputs, making the data distribution more consistent during training and inference.

\subsection{Dynamic corpus}
In the field of information retrieval, databases are evolving entities that expand over time. Unlike Sparse Retrieval, which efficiently incorporates new documents by simply updating existing index structures, and Dense Retrieval techniques, which can easily convert new documents into new embeddings via an encoder and store them, Generative Retrieval faces greater challenges. When new documents are added to the database, it necessitates reindexing the entire corpus, a process that is both time-consuming and computationally intensive. Thus, finding ways to shorten this process becomes crucial for applying generative retrieval in real-world scenarios.

Moreover, ensuring that the model can assimilate new knowledge while retaining memory of old documents is a significant technical challenge. This involves not just the efficiency of model training but also tests the memory capacity of the model. Overall, these challenges point to a common goal: how to effectively manage the dynamic changes of databases under the premise of ensuring model update speed and memory capacity, thereby achieving more flexible and efficient information retrieval.

\citet{kishore2023incdsi} proposed a method that allows for the real-time addition of documents by modifying the original DSI system. They replaced the T5 architecture with a Bert encoder combined with a classification layer to predict the document's unique identifier, adopting atomic ids. This architectural design allows the classification layer to be viewed as a matrix V, where each row represents an embedding of a document. Therefore, during the retrieval process, calculating the inner product between the embedding of a query and that of each document assesses similarity scores. When new documents are added to the system, it essentially amounts to adding a new row to matrix V, introducing a new document embedding. The optimization challenge then becomes finding a way to maximize the similarity score between the new embedding and its corresponding query while ensuring this score is lower than the similarity scores between other documents and their queries. This ensures precise guidance of queries to the most relevant documents.

\citet{mehta2022dsi++} discovered that models face two types of forgetting challenges. Firstly, as the volume of training data increases, documents that were previously accurately retrievable by the model gradually lose their retrievability. Secondly, when new documents are added and fine-tuned, although the new documents are processed effectively, this significantly reduces the processing efficiency for previously trained documents.

To address the first type of forgetting, \citet{mehta2022dsi++} employed Sharpness-Aware Minimization (SAM) technology for improvement. SAM aims to find a "flat" optimal solution, and previous research has shown that models located at flat optima have stronger generalization performance and are better at retaining memory of early learning content in continuous learning tasks.

For the second type of forgetting, \citet{mehta2022dsi++} proposed the use of Generative Memory technology to mitigate the loss of memory for old documents. This method involves training a model capable of generating queries based on previously seen documents, and introducing these generated queries when fine-tuning new documents, thereby helping the model to recall and maintain its processing capabilities for old documents.

\section{Future Directions}
\subsection{Training Methods}
In the rapidly evolving field of Generative Retrieval (GR), the effectiveness of training methods is crucial for optimizing model performance and ensuring robust information retrieval. This section explores the future directions of training methods by focusing on two critical aspects: the strategies for designing document identifiers (DocIDs) and the quality of query generation. Specifically, we provide insights into the potential of learnable DocIDs to capture nuanced document representations and the importance of high-quality data generation in achieving superior retrieval outcomes. Through this exploration, we seek to inspire future research and development in the realm of generative retrieval.
\subsubsection{DocIDs Strategies}
From the introduction in \hyperref[sec:intro-to-gr-indexing]{Indexing}, we understand that indexing is a crucial process in the field of GR, enabling models to memorize the representation of each document. Imagine if the representation of every document were identical; it would be challenging for the model to retrieve the appropriate document for different queries. Therefore, designing accurate document identifiers (DocIDs) is of paramount importance. In the GR field, most document identifiers are static (\citet{tay2022transformer}, \citet{bevilacqua2022autoregressive}, \citet{tang2023semantic} \citet{sun2024learning}), meaning they are defined by predetermined human rules. These identifiers are either sequential numbers (atomic IDs) or semantically structured DocIDs following hierarchical information. However, can these human-defined identifier rules effectively distinguish different documents? This is a key question worth exploring.

Currently, there is emerging work attempting to use learnable document identifiers. For instance, \citet{sun2024learning} employ an additional pre-trained tokenizer model to classify documents more semantically, using token embeddings as identifier information. \citet{wang2023novo} apply attention mechanisms at the n-gram level of documents to learn the overall embedding of each document. 

By leveraging learnable mechanisms, we shift from manually tailored identifier information to allowing models to autonomously learn how to classify documents. This approach holds significant promise for several reasons. Firstly, it can adapt to the evolving nature of language and document content without requiring constant human intervention. Secondly, learnable identifiers can potentially capture subtle semantic nuances that static identifiers might miss, leading to more accurate and contextually relevant retrieval. Future research should focus on refining these learnable identifier mechanisms, exploring ways to integrate dynamic and context-aware identifier generation.

\subsubsection{Query generation quality}
The essence of Generative Retrieval (GR) lies in transforming the traditional Information Retrieval (IR) process, which typically involves two stages (1. query encoding and 2. similarity comparison with different documents), into a single, unified pipeline. The key distinction in this process is the direct generation of document identifiers related to the query using generative models. Building on this foundation, many works have incorporated various generative methods into different phases to enhance each step of the process.

For example, TOME \cite{ren2023tome} augments queries by generating several related paragraphs and corresponding URLs as identifiers. Similarly, DSI++ \cite{mehta2022dsi++} uses generative models to sample several pseudo-queries for documents as a data augmentation method. These approaches have demonstrated superior performance compared to simply training on the original dataset. Given this, we anticipate that future works will increasingly apply generative models to more components to boost the overall performance of generative retrieval.

Under this premise, the quality of data generated by these models becomes crucial. Generative models are renowned for their ability to produce a rich amount of data at a low cost. However, numerous past studies \citep{he2019data, kariluoto2021quality, budach2022effects} have demonstrated that merely increasing data quantity does not directly translate to improved model performance. On the contrary, data quality is the key factor in enhancing model performance.

Therefore, we believe that future research should focus on how to further improve the quality of query generation in the context of generative retrieval. This could involve pre-training models to better understand the context of information retrieval, ensuring that the generated queries are not only abundant but also of high quality. By improving the quality of the generated data, we believe it can significantly boost the performance and reliability of generative retrieval systems.

\subsubsection{Utilizing Large Language Models}
As discussed in Section~\ref{baseline-comparison}, generative retrieval has predominantly utilized encoder-decoder transformers like T5 \cite{raffel2020exploring} and BART \cite{lewis2019bart}.
However, the trend of generative models is shifting towards decoder-only architectures, such as large language models (LLMs).
LLMs, with their vast pre-training on diverse datasets, excel in understanding and generating human-like text, reasoning over complex relationships, following instructions, and adapting to various contexts.
Despite the popularity of LLMs, few have explored using decoder-only architectures for generative retrieval.
Notably, \citet{li2024corpuslm} employed LLaMA-2, \cite{touvron2023llama} an open-source pre-trained LLM, achieving superior performance across various downstream tasks. This suggests the potential advantages of LLMs, likely due to their enhanced reasoning ability and instruction-following capability from extensive pre-training data. The enormous scale of LLMs could enable indexing larger-scale corpora, improving the scalability of generative retrieval. Additionally, the decoder-only architecture could simplify the generative retrieval framework by unifying input and output representations.

\citet{wang2023generative} also applied LLMs in retrieval with LLM2GR, which eliminates additional retrieval models by generating relevant document titles and passages directly from queries. This method, utilizing constrained decoding and a scoring mechanism, demonstrated efficient retrieval.

Nevertheless, some scholars argue that the outputs of pre-trained generative models like LLMs can be unstable, influenced by factors such as GPU settings and prompt design \cite{zakari2024changinggpu}. This variability raises concerns about their reliability as databases. Future research should explore suitable architectures for generative retrieval, assess the reliability of LLMs for this task, and investigate methods to ensure consistent performance, such as constrained decoding algorithms, under varying conditions.

\subsection{Scalability}
\subsubsection{Identifier}
The current research on improving the scalability and updating of documents in Generative retrieval is mainly focused on Atomic identifiers. Since the training process of Atomic identifiers can be simplified to a method similar to Dense retrieval, the characteristics of Dense retrieval can be utilized to enhance the scalability of Generative retrieval. However, in studies using autoregressive generation methods to generate document identifiers, there are relatively fewer breakthroughs in scalability.
\subsubsection{Document adding and deleting}
In dynamic databases, we often need to update outdated documents or adjust the size of the database, which involves deleting infrequently used documents and adding new information. This document editing process can be seen as two main steps: first, deleting old documents, and then adding new ones. Existing research has proven that generative retrieval has certain efficiency and capabilities in adding new documents. However, there is relatively less research on deleting documents, making the exploration of how to effectively delete documents an important and worthwhile topic for in-depth study.
\subsubsection{Indexing cost}
Unlike dense retrieval, which allows for fast indexing of new documents using a pre-trained encoder, generative retrieval faces a significant challenge in this regard. In generative retrieval, indexing a new document often necessitates retraining the entire model, a process that is both time-consuming and resource-intensive. This bottleneck in the indexing process can hinder the practical application of generative retrieval systems in dynamic environments where documents are frequently updated. Therefore, exploring alternative methods to approximate the indexing of new documents without requiring full model retraining presents a promising direction for future research. Such methods could potentially bridge the gap in indexing efficiency between dense and generative retrieval approaches, enhancing the practicality and scalability of generative retrieval systems

\subsection{Multi-task}

In the framework of seq2seq architecture within the field of Information Retrieval (IR), generative retrieval, operating on a single-task basis, not only outperforms the sparse retrieval baseline but also demonstrates substantial enhancements in Zero-Shot NQ document retrieval \cite{tay2022transformer}. The authors of the DSI have employed a T5-style co-training approach, implementing simultaneous indexing and retrieval using different task prompts. However, subsequent studies have yet to experiment with this methodology in tasks related to Question Answering, summarization, or Semantic Textual Similarity. Integrating downstream tasks directly into the IR tasks using T5-style co-training presents a promising future direction.

Another direction could involve enhancing the capabilities of generative retrieval and downstream tasks within an architecture similar to DSI by incorporating additional components. For instance, the UniGen \cite{li2024unigen} has incorporated a Question Answering task, utilizing a shared encoder and employing two distinct decoders to efficiently execute IR and QA tasks, achieving notable results. Alternatively, integrating the Retrieval-Augmented Generation (RAG) architecture \cite{li2024towards} could facilitate the co-training of multiple downstream tasks beyond the IR task. Additionally, employing a Mixture of Experts could augment the capabilities of various downstream tasks, achieving enhanced overall performance through gating mechanisms \cite{shazeer2017outrageously} that control different modules.

\section{Conclusion}
This paper provides a comprehensive survey and analysis of Generative Retrieval (GR), exploring its development history, key technologies, challenges, and future directions. Five significant contributions to the field of information retrieval are provided in the following:
\begin{itemize}
\item The development of information retrieval is traced from sparse retrieval methods to dense retrieval techniques, eventually leading to generative retrieval, wherein a query is directly mapped to its relevant document(s) through a seq2seq model without the need for pre-retrieval query processing or post-retrieval reranking of documents.
\item The core concepts of GR are explained, detailing the end-to-end retrieval process, indexing, and retrieval techniques, including document identifier strategies and seq2seq models.
\item Various document identifier types are compared, showing that those with semantic information generally perform better, and different methods of creating these identifiers are explored.
\item Evaluation metrics and commonly used datasets in GR are discussed, emphasizing their role in assessing retrieval performance and comparing different identifier strategies.
\item Challenges such as scalability and dynamic corpus management are identified. Future research directions are proposed, such as optimizing training methods, improving system scalability, and integrating multi-task learning techniques.
\end{itemize}
In summary, this study offers a thorough survey to help readers gain a deeper understanding of generative retrieval technologies. It aims to inspire further research in the field and advance the development of information retrieval technologies.

\section*{Acknowledgements}
This work was financially supported by the National Science and Technology Council (NSTC) in Taiwan, under Grants 112-2223-E-002-012-MY5 and 111-2222-E-002-013-MY3, and from the Featured Area Research Center Program within the framework of the Higher Education Sprout Project by the Ministry of Education (113L900901/113L900902/113L900903).

\bibliography{custom}

\begin{thebibliography}{57}
\expandafter\ifx\csname natexlab\endcsname\relax\def\natexlab#1{#1}\fi

\bibitem[{Asai et~al.(2020)Asai, Kasai, Clark, Lee, Choi, and Hajishirzi}]{asai2020xor}
Akari Asai, Jungo Kasai, Jonathan~H Clark, Kenton Lee, Eunsol Choi, and Hannaneh Hajishirzi. 2020.
\newblock Xor qa: Cross-lingual open-retrieval question answering.
\newblock \emph{arXiv preprint arXiv:2010.11856}.

\bibitem[{Bevilacqua et~al.(2022)Bevilacqua, Ottaviano, Lewis, Yih, Riedel, and Petroni}]{bevilacqua2022autoregressive}
Michele Bevilacqua, Giuseppe Ottaviano, Patrick Lewis, Scott Yih, Sebastian Riedel, and Fabio Petroni. 2022.
\newblock Autoregressive search engines: Generating substrings as document identifiers.
\newblock \emph{Advances in Neural Information Processing Systems}, 35:31668--31683.

\bibitem[{Budach et~al.(2022)Budach, Feuerpfeil, Ihde, Nathansen, Noack, Patzlaff, Naumann, and Harmouch}]{budach2022effects}
Lukas Budach, Moritz Feuerpfeil, Nina Ihde, Andrea Nathansen, Nele Noack, Hendrik Patzlaff, Felix Naumann, and Hazar Harmouch. 2022.
\newblock \href {http://arxiv.org/abs/2207.14529} {The effects of data quality on machine learning performance}.

\bibitem[{Chen et~al.(2023)Chen, Zhang, Guo, de~Rijke, Chen, Fan, and Cheng}]{chen2023continual}
Jiangui Chen, Ruqing Zhang, Jiafeng Guo, Maarten de~Rijke, Wei Chen, Yixing Fan, and Xueqi Cheng. 2023.
\newblock Continual learning for generative retrieval over dynamic corpora.
\newblock In \emph{Proceedings of the 32nd ACM International Conference on Information and Knowledge Management}, pages 306--315.

\bibitem[{Chen et~al.(2022{\natexlab{a}})Chen, Zhang, Guo, Fan, and Cheng}]{chen2022gere}
Jiangui Chen, Ruqing Zhang, Jiafeng Guo, Yixing Fan, and Xueqi Cheng. 2022{\natexlab{a}}.
\newblock Gere: Generative evidence retrieval for fact verification.
\newblock In \emph{Proceedings of the 45th International ACM SIGIR Conference on Research and Development in Information Retrieval}, pages 2184--2189.

\bibitem[{Chen et~al.(2022{\natexlab{b}})Chen, Zhang, Guo, Liu, Fan, and Cheng}]{chen2022corpusbrain}
Jiangui Chen, Ruqing Zhang, Jiafeng Guo, Yiqun Liu, Yixing Fan, and Xueqi Cheng. 2022{\natexlab{b}}.
\newblock Corpusbrain: Pre-train a generative retrieval model for knowledge-intensive language tasks.
\newblock In \emph{Proceedings of the 31st ACM International Conference on Information \& Knowledge Management}, pages 191--200.

\bibitem[{De~Cao et~al.(2020)De~Cao, Izacard, Riedel, and Petroni}]{de2020autoregressive}
Nicola De~Cao, Gautier Izacard, Sebastian Riedel, and Fabio Petroni. 2020.
\newblock Autoregressive entity retrieval.
\newblock \emph{arXiv preprint arXiv:2010.00904}.

\bibitem[{Devlin et~al.(2018)Devlin, Chang, Lee, and Toutanova}]{devlin2018bert}
Jacob Devlin, Ming-Wei Chang, Kenton Lee, and Kristina Toutanova. 2018.
\newblock Bert: Pre-training of deep bidirectional transformers for language understanding.
\newblock \emph{arXiv preprint arXiv:1810.04805}.

\bibitem[{Guo et~al.(2016)Guo, Fan, Ai, and Croft}]{guo2016deep}
Jiafeng Guo, Yixing Fan, Qingyao Ai, and W~Bruce Croft. 2016.
\newblock A deep relevance matching model for ad-hoc retrieval.
\newblock In \emph{Proceedings of the 25th ACM international on conference on information and knowledge management}, pages 55--64.

\bibitem[{Guu et~al.(2020)Guu, Lee, Tung, Pasupat, and Chang}]{guu2020retrieval}
Kelvin Guu, Kenton Lee, Zora Tung, Panupong Pasupat, and Mingwei Chang. 2020.
\newblock Retrieval augmented language model pre-training.
\newblock In \emph{International conference on machine learning}, pages 3929--3938. PMLR.

\bibitem[{He et~al.(2019)He, Yu, Wang, Li, and Chen}]{he2019data}
Tianxing He, Shengcheng Yu, Ziyuan Wang, Jieqiong Li, and Zhenyu Chen. 2019.
\newblock \href {http://arxiv.org/abs/1906.11882} {From data quality to model quality: an exploratory study on deep learning}.

\bibitem[{Husain et~al.(2019)Husain, Wu, Gazit, Allamanis, and Brockschmidt}]{husain2019codesearchnet}
Hamel Husain, Ho-Hsiang Wu, Tiferet Gazit, Miltiadis Allamanis, and Marc Brockschmidt. 2019.
\newblock Codesearchnet challenge: Evaluating the state of semantic code search.
\newblock \emph{arXiv preprint arXiv:1909.09436}.

\bibitem[{Jiang et~al.(2020)Jiang, Bordia, Zhong, Dognin, Singh, and Bansal}]{jiang2020hover}
Yichen Jiang, Shikha Bordia, Zheng Zhong, Charles Dognin, Maneesh Singh, and Mohit Bansal. 2020.
\newblock Hover: A dataset for many-hop fact extraction and claim verification.
\newblock \emph{arXiv preprint arXiv:2011.03088}.

\bibitem[{Joshi et~al.(2017)Joshi, Choi, Weld, and Zettlemoyer}]{joshi2017triviaqa}
Mandar Joshi, Eunsol Choi, Daniel~S Weld, and Luke Zettlemoyer. 2017.
\newblock Triviaqa: A large scale distantly supervised challenge dataset for reading comprehension.
\newblock \emph{arXiv preprint arXiv:1705.03551}.

\bibitem[{Kariluoto et~al.(2021)Kariluoto, Pärnänen, Kultanen, Soininen, and Abrahamsson}]{kariluoto2021quality}
Antti Kariluoto, Arto Pärnänen, Joni Kultanen, Jukka Soininen, and Pekka Abrahamsson. 2021.
\newblock \href {http://arxiv.org/abs/2112.09400} {Quality of data in machine learning}.

\bibitem[{Karpukhin et~al.(2020)Karpukhin, O{\u{g}}uz, Min, Lewis, Wu, Edunov, Chen, and Yih}]{karpukhin2020dense}
Vladimir Karpukhin, Barlas O{\u{g}}uz, Sewon Min, Patrick Lewis, Ledell Wu, Sergey Edunov, Danqi Chen, and Wen-tau Yih. 2020.
\newblock Dense passage retrieval for open-domain question answering.
\newblock \emph{arXiv preprint arXiv:2004.04906}.

\bibitem[{Khattab et~al.(2021)Khattab, Potts, and Zaharia}]{khattab2021relevance}
Omar Khattab, Christopher Potts, and Matei Zaharia. 2021.
\newblock Relevance-guided supervision for openqa with colbert.
\newblock \emph{Transactions of the association for computational linguistics}, 9:929--944.

\bibitem[{Kishore et~al.(2023)Kishore, Wan, Lovelace, Artzi, and Weinberger}]{kishore2023incdsi}
Varsha Kishore, Chao Wan, Justin Lovelace, Yoav Artzi, and Kilian~Q Weinberger. 2023.
\newblock Incdsi: incrementally updatable document retrieval.
\newblock In \emph{International Conference on Machine Learning}, pages 17122--17134. PMLR.

\bibitem[{Kwiatkowski et~al.(2019)Kwiatkowski, Palomaki, Redfield, Collins, Parikh, Alberti, Epstein, Polosukhin, Devlin, Lee et~al.}]{kwiatkowski2019natural}
Tom Kwiatkowski, Jennimaria Palomaki, Olivia Redfield, Michael Collins, Ankur Parikh, Chris Alberti, Danielle Epstein, Illia Polosukhin, Jacob Devlin, Kenton Lee, et~al. 2019.
\newblock Natural questions: a benchmark for question answering research.
\newblock \emph{Transactions of the Association for Computational Linguistics}, 7:453--466.

\bibitem[{Lewis et~al.(2019)Lewis, Liu, Goyal, Ghazvininejad, Mohamed, Levy, Stoyanov, and Zettlemoyer}]{lewis2019bart}
Mike Lewis, Yinhan Liu, Naman Goyal, Marjan Ghazvininejad, Abdelrahman Mohamed, Omer Levy, Ves Stoyanov, and Luke Zettlemoyer. 2019.
\newblock Bart: Denoising sequence-to-sequence pre-training for natural language generation, translation, and comprehension.
\newblock \emph{arXiv preprint arXiv:1910.13461}.

\bibitem[{Lewis et~al.(2020)Lewis, Perez, Piktus, Petroni, Karpukhin, Goyal, K{\"u}ttler, Lewis, Yih, Rockt{\"a}schel et~al.}]{lewis2020retrieval}
Patrick Lewis, Ethan Perez, Aleksandra Piktus, Fabio Petroni, Vladimir Karpukhin, Naman Goyal, Heinrich K{\"u}ttler, Mike Lewis, Wen-tau Yih, Tim Rockt{\"a}schel, et~al. 2020.
\newblock Retrieval-augmented generation for knowledge-intensive nlp tasks.
\newblock \emph{Advances in Neural Information Processing Systems}, 33:9459--9474.

\bibitem[{Li et~al.(2024{\natexlab{a}})Li, Dou, Zhou, and Liu}]{li2024corpuslm}
Xiaoxi Li, Zhicheng Dou, Yujia Zhou, and Fangchao Liu. 2024{\natexlab{a}}.
\newblock Corpuslm: Towards a unified language model on corpus for knowledge-intensive tasks.

\bibitem[{Li et~al.(2024{\natexlab{b}})Li, Dou, Zhou, and Liu}]{li2024towards}
Xiaoxi Li, Zhicheng Dou, Yujia Zhou, and Fangchao Liu. 2024{\natexlab{b}}.
\newblock Towards a unified language model for knowledge-intensive tasks utilizing external corpus.
\newblock \emph{arXiv preprint arXiv:2402.01176}.

\bibitem[{Li et~al.(2024{\natexlab{c}})Li, Zhou, and Dou}]{li2024unigen}
Xiaoxi Li, Yujia Zhou, and Zhicheng Dou. 2024{\natexlab{c}}.
\newblock Unigen: A unified generative framework for retrieval and question answering with large language models.
\newblock In \emph{Proceedings of the AAAI Conference on Artificial Intelligence}, volume~38, pages 8688--8696.

\bibitem[{Li et~al.(2023)Li, Yang, Wang, Wei, and Li}]{li2023multiview}
Yongqi Li, Nan Yang, Liang Wang, Furu Wei, and Wenjie Li. 2023.
\newblock Multiview identifiers enhanced generative retrieval.
\newblock \emph{arXiv preprint arXiv:2305.16675}.

\bibitem[{Li et~al.(2024{\natexlab{d}})Li, Yang, Wang, Wei, and Li}]{li2024learning}
Yongqi Li, Nan Yang, Liang Wang, Furu Wei, and Wenjie Li. 2024{\natexlab{d}}.
\newblock Learning to rank in generative retrieval.
\newblock In \emph{Proceedings of the AAAI Conference on Artificial Intelligence}, volume~38, pages 8716--8723.

\bibitem[{Mehta et~al.(2022)Mehta, Gupta, Tay, Dehghani, Tran, Rao, Najork, Strubell, and Metzler}]{mehta2022dsi++}
Sanket~Vaibhav Mehta, Jai Gupta, Yi~Tay, Mostafa Dehghani, Vinh~Q Tran, Jinfeng Rao, Marc Najork, Emma Strubell, and Donald Metzler. 2022.
\newblock Dsi++: Updating transformer memory with new documents.
\newblock \emph{arXiv preprint arXiv:2212.09744}.

\bibitem[{Mikolov et~al.(2013)Mikolov, Chen, Corrado, and Dean}]{mikolov2013efficient}
Tomas Mikolov, Kai Chen, Greg Corrado, and Jeffrey Dean. 2013.
\newblock Efficient estimation of word representations in vector space.
\newblock \emph{arXiv preprint arXiv:1301.3781}.

\bibitem[{Mitra et~al.(2017)Mitra, Diaz, and Craswell}]{mitra2017learning}
Bhaskar Mitra, Fernando Diaz, and Nick Craswell. 2017.
\newblock Learning to match using local and distributed representations of text for web search.
\newblock In \emph{Proceedings of the 26th international conference on world wide web}, pages 1291--1299.

\bibitem[{Nadeem et~al.(2022)Nadeem, Ziems, and Wu}]{nadeem2022codedsi}
Usama Nadeem, Noah Ziems, and Shaoen Wu. 2022.
\newblock Codedsi: Differentiable code search.
\newblock \emph{arXiv preprint arXiv:2210.00328}.

\bibitem[{Nguyen and Yates(2023)}]{nguyen2023generative}
Thong Nguyen and Andrew Yates. 2023.
\newblock Generative retrieval as dense retrieval.
\newblock \emph{arXiv preprint arXiv:2306.11397}.

\bibitem[{Nguyen et~al.(2016)Nguyen, Rosenberg, Song, Gao, Tiwary, Majumder, and Deng}]{nguyen2016ms}
Tri Nguyen, Mir Rosenberg, Xia Song, Jianfeng Gao, Saurabh Tiwary, Rangan Majumder, and Li~Deng. 2016.
\newblock Ms marco: A human-generated machine reading comprehension dataset.

\bibitem[{Nogueira et~al.(2019)Nogueira, Yang, Lin, and Cho}]{nogueira2019document}
Rodrigo Nogueira, Wei Yang, Jimmy Lin, and Kyunghyun Cho. 2019.
\newblock Document expansion by query prediction.
\newblock \emph{arXiv preprint arXiv:1904.08375}.

\bibitem[{Pennington et~al.(2014)Pennington, Socher, and Manning}]{pennington2014glove}
Jeffrey Pennington, Richard Socher, and Christopher~D Manning. 2014.
\newblock Glove: Global vectors for word representation.
\newblock In \emph{Proceedings of the 2014 conference on empirical methods in natural language processing (EMNLP)}, pages 1532--1543.

\bibitem[{Petroni et~al.(2020)Petroni, Piktus, Fan, Lewis, Yazdani, De~Cao, Thorne, Jernite, Karpukhin, Maillard et~al.}]{petroni2020kilt}
Fabio Petroni, Aleksandra Piktus, Angela Fan, Patrick Lewis, Majid Yazdani, Nicola De~Cao, James Thorne, Yacine Jernite, Vladimir Karpukhin, Jean Maillard, et~al. 2020.
\newblock Kilt: a benchmark for knowledge intensive language tasks.
\newblock \emph{arXiv preprint arXiv:2009.02252}.

\bibitem[{Pradeep et~al.(2023)Pradeep, Hui, Gupta, Lelkes, Zhuang, Lin, Metzler, and Tran}]{pradeep2023does}
Ronak Pradeep, Kai Hui, Jai Gupta, Adam~D Lelkes, Honglei Zhuang, Jimmy Lin, Donald Metzler, and Vinh~Q Tran. 2023.
\newblock How does generative retrieval scale to millions of passages?
\newblock \emph{arXiv preprint arXiv:2305.11841}.

\bibitem[{Raffel et~al.(2020)Raffel, Shazeer, Roberts, Lee, Narang, Matena, Zhou, Li, and Liu}]{raffel2020exploring}
Colin Raffel, Noam Shazeer, Adam Roberts, Katherine Lee, Sharan Narang, Michael Matena, Yanqi Zhou, Wei Li, and Peter~J Liu. 2020.
\newblock Exploring the limits of transfer learning with a unified text-to-text transformer.
\newblock \emph{Journal of machine learning research}, 21(140):1--67.

\bibitem[{Ren et~al.(2023)Ren, Zhao, Liu, Wu, Wen, and Wang}]{ren2023tome}
Ruiyang Ren, Wayne~Xin Zhao, Jing Liu, Hua Wu, Ji-Rong Wen, and Haifeng Wang. 2023.
\newblock Tome: A two-stage approach for model-based retrieval.
\newblock \emph{arXiv preprint arXiv:2305.11161}.

\bibitem[{Robertson and Jones(1976)}]{robertson1976relevance}
Stephen~E Robertson and K~Sparck Jones. 1976.
\newblock Relevance weighting of search terms.
\newblock \emph{Journal of the American Society for Information science}, 27(3):129--146.

\bibitem[{Salton(1983)}]{salton1983introduction}
Gerard Salton. 1983.
\newblock Introduction to modern information retrieval.
\newblock \emph{McGraw-Hill}.

\bibitem[{Shazeer et~al.(2017)Shazeer, Mirhoseini, Maziarz, Davis, Le, Hinton, and Dean}]{shazeer2017outrageously}
Noam Shazeer, Azalia Mirhoseini, Krzysztof Maziarz, Andy Davis, Quoc Le, Geoffrey Hinton, and Jeff Dean. 2017.
\newblock Outrageously large neural networks: The sparsely-gated mixture-of-experts layer.
\newblock \emph{arXiv preprint arXiv:1701.06538}.

\bibitem[{Sun et~al.(2024)Sun, Yan, Chen, Wang, Zhu, Ren, Chen, Yin, Rijke, and Ren}]{sun2024learning}
Weiwei Sun, Lingyong Yan, Zheng Chen, Shuaiqiang Wang, Haichao Zhu, Pengjie Ren, Zhumin Chen, Dawei Yin, Maarten Rijke, and Zhaochun Ren. 2024.
\newblock Learning to tokenize for generative retrieval.
\newblock \emph{Advances in Neural Information Processing Systems}, 36.

\bibitem[{Tang et~al.(2023{\natexlab{a}})Tang, Zhang, Guo, Chen, Zhu, Wang, Yin, and Cheng}]{tang2023semantic}
Yubao Tang, Ruqing Zhang, Jiafeng Guo, Jiangui Chen, Zuowei Zhu, Shuaiqiang Wang, Dawei Yin, and Xueqi Cheng. 2023{\natexlab{a}}.
\newblock Semantic-enhanced differentiable search index inspired by learning strategies.
\newblock In \emph{Proceedings of the 29th ACM SIGKDD Conference on Knowledge Discovery and Data Mining}, pages 4904--4913.

\bibitem[{Tang et~al.(2023{\natexlab{b}})Tang, Zhang, Guo, and de~Rijke}]{tang2023recent}
Yubao Tang, Ruqing Zhang, Jiafeng Guo, and Maarten de~Rijke. 2023{\natexlab{b}}.
\newblock Recent advances in generative information retrieval.
\newblock In \emph{Proceedings of the Annual International ACM SIGIR Conference on Research and Development in Information Retrieval in the Asia Pacific Region}, pages 294--297.

\bibitem[{Tay et~al.(2022)Tay, Tran, Dehghani, Ni, Bahri, Mehta, Qin, Hui, Zhao, Gupta et~al.}]{tay2022transformer}
Yi~Tay, Vinh Tran, Mostafa Dehghani, Jianmo Ni, Dara Bahri, Harsh Mehta, Zhen Qin, Kai Hui, Zhe Zhao, Jai Gupta, et~al. 2022.
\newblock Transformer memory as a differentiable search index.
\newblock \emph{Advances in Neural Information Processing Systems}, 35:21831--21843.

\bibitem[{Thorne(2022)}]{thorne2022data}
James Thorne. 2022.
\newblock Data-efficient autoregressive document retrieval for fact verification.
\newblock \emph{arXiv preprint arXiv:2211.09388}.

\bibitem[{Thorne et~al.(2018)Thorne, Vlachos, Christodoulopoulos, and Mittal}]{thorne2018fever}
James Thorne, Andreas Vlachos, Christos Christodoulopoulos, and Arpit Mittal. 2018.
\newblock Fever: a large-scale dataset for fact extraction and verification.
\newblock \emph{arXiv preprint arXiv:1803.05355}.

\bibitem[{Touvron et~al.(2023)Touvron, Martin, Stone, Albert, Almahairi, Babaei, Bashlykov, Batra, Bhargava, Bhosale et~al.}]{touvron2023llama}
Hugo Touvron, Louis Martin, Kevin Stone, Peter Albert, Amjad Almahairi, Yasmine Babaei, Nikolay Bashlykov, Soumya Batra, Prajjwal Bhargava, Shruti Bhosale, et~al. 2023.
\newblock Llama 2: Open foundation and fine-tuned chat models.
\newblock \emph{arXiv preprint arXiv:2307.09288}.

\bibitem[{Wang et~al.(2023{\natexlab{a}})Wang, Xie, Hu, Ye, and Zhang}]{wang2023generative}
Ye~Wang, Rui Xie, Wenxin Hu, Wei Ye, and Shikun Zhang. 2023{\natexlab{a}}.
\newblock Generative retrieval with large language models.

\bibitem[{Wang et~al.(2022)Wang, Hou, Wang, Miao, Wu, Chen, Xia, Chi, Zhao, Liu et~al.}]{wang2022neural}
Yujing Wang, Yingyan Hou, Haonan Wang, Ziming Miao, Shibin Wu, Qi~Chen, Yuqing Xia, Chengmin Chi, Guoshuai Zhao, Zheng Liu, et~al. 2022.
\newblock A neural corpus indexer for document retrieval.
\newblock \emph{Advances in Neural Information Processing Systems}, 35:25600--25614.

\bibitem[{Wang et~al.(2023{\natexlab{b}})Wang, Zhou, Tu, and Dou}]{wang2023novo}
Zihan Wang, Yujia Zhou, Yiteng Tu, and Zhicheng Dou. 2023{\natexlab{b}}.
\newblock Novo: Learnable and interpretable document identifiers for model-based ir.
\newblock In \emph{Proceedings of the 32nd ACM International Conference on Information and Knowledge Management}, pages 2656--2665.

\bibitem[{Xiao et~al.(2022)Xiao, Liu, Shao, and Cao}]{xiao2022retromae}
Shitao Xiao, Zheng Liu, Yingxia Shao, and Zhao Cao. 2022.
\newblock \href {http://arxiv.org/abs/2205.12035} {Retromae: Pre-training retrieval-oriented language models via masked auto-encoder}.

\bibitem[{Zakari(2024)}]{zakari2024changinggpu}
Anis Zakari. 2024.
\newblock Changing the gpu is changing the behaviour of your llm.
\newblock \url{https://medium.com/@anis.zakari/changing-the-gpu-is-changing-the-behaviour-of-your-llm-0e6dd8dfaaae}.
\newblock Accessed: 2024-06-02.

\bibitem[{Zhang et~al.(2023)Zhang, Liu, Zhou, Dou, and Cao}]{zhang2023term}
Peitian Zhang, Zheng Liu, Yujia Zhou, Zhicheng Dou, and Zhao Cao. 2023.
\newblock Term-sets can be strong document identifiers for auto-regressive search engines.
\newblock \emph{arXiv preprint arXiv:2305.13859}.

\bibitem[{Zhou et~al.(2022{\natexlab{a}})Zhou, Yao, Dou, Wu, and Wen}]{zhou2022dynamicretriever}
Yujia Zhou, Jing Yao, Zhicheng Dou, Ledell Wu, and Ji-Rong Wen. 2022{\natexlab{a}}.
\newblock Dynamicretriever: A pre-training model-based ir system with neither sparse nor dense index.
\newblock \emph{arXiv preprint arXiv:2203.00537}.

\bibitem[{Zhou et~al.(2022{\natexlab{b}})Zhou, Yao, Dou, Wu, Zhang, and Wen}]{zhou2022ultron}
Yujia Zhou, Jing Yao, Zhicheng Dou, Ledell Wu, Peitian Zhang, and Ji-Rong Wen. 2022{\natexlab{b}}.
\newblock Ultron: An ultimate retriever on corpus with a model-based indexer.
\newblock \emph{arXiv preprint arXiv:2208.09257}.

\bibitem[{Zhuang et~al.(2022)Zhuang, Ren, Shou, Pei, Gong, Zuccon, and Jiang}]{zhuang2022bridging}
Shengyao Zhuang, Houxing Ren, Linjun Shou, Jian Pei, Ming Gong, Guido Zuccon, and Daxin Jiang. 2022.
\newblock Bridging the gap between indexing and retrieval for differentiable search index with query generation.
\newblock \emph{arXiv preprint arXiv:2206.10128}.

\end{thebibliography}

\appendix

\end{document}